\documentclass[showpacs, oneside, twocolumn, prd, amsmath, amssymb, nofootinbib, superscriptaddress]{revtex4-1}

\usepackage{cases}
\usepackage{amsmath}
\usepackage{amssymb}
\usepackage{amsfonts}
\usepackage{amssymb}
\usepackage{dcolumn}
\usepackage{bm}
\usepackage{bbm}
\usepackage{graphicx}
\usepackage{xcolor}
\usepackage{array}
\usepackage{subfigure}
\usepackage{hyperref}
\usepackage{wasysym}
\usepackage{mathrsfs}

\newcommand{\be}{\begin{equation}}
\newcommand{\ee}{\end{equation}}
\newcommand{\ba}{\begin{eqnarray}}
\newcommand{\ea}{\end{eqnarray}}

\newcommand{\gsim}{\mathrel{\hbox{\rlap{\lower.55ex \hbox {$\sim$}}
                   \kern-.3em \raise.4ex \hbox{$>$}}}}
\newcommand{\lsim}{\mathrel{\hbox{\rlap{\lower.55ex \hbox {$\sim$}}
                   \kern-.3em \raise.4ex \hbox{$<$}}}}

\hypersetup{colorlinks=true,
            breaklinks=true,
            pdfstartview=Fit,
            linkcolor=blue,
            citecolor=blue,
            urlcolor=blue}

\begin{document}
\title{Testing the equivalence principle via the shadow of black holes}

\author{Sheng-Feng Yan}
\affiliation{Department of Astronomy, School of Physical Sciences, University of Science and Technology of China, Hefei, Anhui 230026, China}
\affiliation{CAS Key Laboratory for Research in Galaxies and Cosmology, University of Science and Technology of China, Hefei, Anhui 230026, China}
\affiliation{School of Astronomy and Space Science, University of Science and Technology of China, Hefei, Anhui 230026, China}

\author{Chunlong Li}
\affiliation{Department of Astronomy, School of Physical Sciences, University of Science and Technology of China, Hefei, Anhui 230026, China}
\affiliation{CAS Key Laboratory for Research in Galaxies and Cosmology, University of Science and Technology of China, Hefei, Anhui 230026, China}
\affiliation{School of Astronomy and Space Science, University of Science and Technology of China, Hefei, Anhui 230026, China}

\author{Lingqin Xue}
\affiliation{Department of Astronomy, School of Physical Sciences, University of Science and Technology of China, Hefei, Anhui 230026, China}
\affiliation{CAS Key Laboratory for Research in Galaxies and Cosmology, University of Science and Technology of China, Hefei, Anhui 230026, China}
\affiliation{School of Astronomy and Space Science, University of Science and Technology of China, Hefei, Anhui 230026, China}
\affiliation{Department of Physics, University of Florida, Gainesville, FL 32611-8440, United States of America}

\author{Xin Ren}
\affiliation{Department of Astronomy, School of Physical Sciences, University of Science and Technology of China, Hefei, Anhui 230026, China}
\affiliation{CAS Key Laboratory for Research in Galaxies and Cosmology, University of Science and Technology of China, Hefei, Anhui 230026, China}
\affiliation{School of Astronomy and Space Science, University of Science and Technology of China, Hefei, Anhui 230026, China}

\author{Yi-Fu Cai}
\email{yifucai@ustc.edu.cn}
\affiliation{Department of Astronomy, School of Physical Sciences, University of Science and Technology of China, Hefei, Anhui 230026, China}
\affiliation{CAS Key Laboratory for Research in Galaxies and Cosmology, University of Science and Technology of China, Hefei, Anhui 230026, China}
\affiliation{School of Astronomy and Space Science, University of Science and Technology of China, Hefei, Anhui 230026, China}

\author{Damien A. Easson}
\email{easson@asu.edu}
\affiliation{Department of Physics, Arizona State University, Tempe, AZ 85287-1504, USA}

\author{Ye-Fei Yuan}
\email{yfyuan@ustc.edu.cn}
\affiliation{Department of Astronomy, School of Physical Sciences, University of Science and Technology of China, Hefei, Anhui 230026, China}
\affiliation{CAS Key Laboratory for Research in Galaxies and Cosmology, University of Science and Technology of China, Hefei, Anhui 230026, China}
\affiliation{School of Astronomy and Space Science, University of Science and Technology of China, Hefei, Anhui 230026, China}

\author{Hongsheng Zhao}
\email{hz4@st-andrews.ac.uk}
\affiliation{Department of Astronomy, School of Physical Sciences, University of Science and Technology of China, Hefei, Anhui 230026, China}
\affiliation{Scottish Universities Physics Alliance, University of St Andrews, North Haugh, St Andrews, Fife KY16 9SS, UK}

\begin{abstract}

We study the equivalence principle, regarded as the cornerstone of general relativity, by analyzing the deformation observable of black hole shadows. Such deformation can arise from new physics and may be expressed as a phenomenological violation of the equivalence principle. Specifically, we assume that there is an additional background vector field 
that couples to 
the photons around the supermassive black hole. This type of coupling yields impact on the way the system depends on initial conditions, and affects the black hole shadow at different wavelengths by a different amount, and therefore observations of the shadow in different wavelengths could constrain such couplings. 
This can be tested by future multi-band observations. Adopting a specific form of the vector field, we obtain constraints on model parameters from Event Horizon Telescope observations and measurements of gas/stellar orbits.
%

\end{abstract}

\pacs{04.70.-s, 97.60.Lf, 11.30.Cp, 98.80.Cq}


\maketitle

\section{Introduction}

The first image of the supermassive black hole (SMBH) M87 observed by the Event Horizon Telescope (EHT) \cite{Akiyama:2019cqa, Akiyama:2019brx, Akiyama:2019sww, Akiyama:2019bqs, Akiyama:2019fyp, Akiyama:2019eap} led us into a new era of black hole physics. The high spatial resolution makes direct visual observation of a SMBH and surrounding environment possible. With this way of studying the most extreme objects in our universe predicted by Einstein's theory of General Relativity (GR),  scientists can explore the nature of fundamental physics from the information we obtain from the EHT and other forthcoming experiments.

The black hole photograph taken by the EHT project provides us with direct information about the motion of the photons near the event horizon scale for the first time. Within standard GR, the trajectories of test photons correspond to geodesic paths, which are fully determined by the metric tensor field. Any new physics that could be revealed by the detection of the photon's path should have modifications to the standard expression of the metric. This principle has spawned a series of works in the literature \cite{Amorim:2019xrp, Cardoso:2019rvt, Nokhrina:2019sxv, Wang:2019tjc, Kapec:2019hro, Jusufi:2019nrn, Contreras:2019cmf, Konoplya:2019goy, Davies:2019wgi, Qi:2019zdk, Tsupko:2019pzg, Firouzjaee:2019aij, Tsupko:2019mfo, Rezzolla:2014mua, Allahyari:2019jqz, Konoplya:2016jvv, Giddings:2016btb, Giddings:2019jwy, Bambi:2019tjh}. For example, the violation of rigid vacuum solutions of the black hole metric due to the accumulation of extra mass around the black hole could leave observable effects on the black shadow, such as the formation of ultralight boson clouds around Kerr black holes caused by the super-radiance process \cite{Roy:2019esk, Bar:2019pnz, Davoudiasl:2019nlo, Cunha:2019ikd}. Moreover, gravitational theories beyond GR that lead to different black hole metrics can also affect the shape of black hole shadows. Examples could be found in theories of asymptotically safe gravity motivated by the renormalization issues of quantum gravity \cite{Held:2019xde, Kumar:2019ohr, Cai:2010zh}, and other classical modifications of GR \cite{Izmailov:2019uhy, Ovgun:2019jdo, Zhu:2019ura, Tian:2019yhn, Amarilla:2015pgp, Vagnozzi:2019apd, Long:2019nox, Banerjee:2019nnj}. Some non-singular black hole solutions outside of GR may also lead to additional features of the shape and size of the shadow \cite{Tsukamoto:2017fxq, Abdujabbarov:2016hnw, Dymnikova:2019vuz, Lamy:2018zvj, Kumar:2019pjp}.

In addition, there are some theories that are beyond the regular metric description of black hole spacetimes that have impacts on the motion of photons. One such case is when the gravity theory differs from  standard Riemannian geometry. For instance, an additional torsional tensor present in gravity theories based on the Riemann-Cartan spacetime \cite{Hehl:1976kj}, where the spacetime torsion can behave as a force causing the motion of particles to deviate from the usual extremal paths predicted by GR \cite{Prasanna:2009zz, Wanas:1999if, Yasskin:1980bu, Novello:1976tx, DeSabbata:1980gb, Hayashi:1980av, Kleinert:1998xa, Hammond:2018man, Poplawski:2011cr, Cai:2015emx, Cai:2018rzd, Li:2018ixg, Chen:2019ftv, Yan:2019gbw}. Another case arises in particle physics when the super-radiance process of a rotating black hole may extract mass and angular momentum from the black hole to produce an accumulation of light bosons to form a macroscopic ``cloud" composed of a  bosonic field condensate~\cite{ZelDovich:1972, Detweiler:1980uk, Nielsen:2019izz, Baumann:2019eav, Huang:2019xbu}. Such light bosons can come from physics beyond the standard model such as  axions or light gauge bosons of hidden $U(1)$ symmetries \cite{Pawl:2004bx, Arvanitaki:2014wva, Chen:2019fsq}. If these particles have weak couplings to the photons, then photon paths can be affected by leaving observable effects on the black hole shadow. These facts inspire us to explore phenomena that might be caused by new physics that is not caused simply by modifications of the metric. 

In this paper, we propose to use the shadow of SMBH as a probe to detect underlying new physics whose effects on photon motion cannot be described by the metric and could, phenomenologically, be regarded as a violation of the equivalence principle. Specifically, we consider an additional background vector field effectively generated by the central black hole. This background field couples to photons and behaves as a ``force" acting on the photon via a coupling constant. The effect is analogous to the motion of charged particles in an electromagnetic field generated by a charged black hole. 
Phenomenologically, this could be regarded as a violation of the weak equivalence principle \cite{Will:2014kxa, Tino:2020nla}.

To extract model-independent results, we do not start from a specific theory of the background vector field. Instead, given that the vector field and the metric field are generated by the same source, it is physically reasonable to assume the vector field has the same symmetry as the space-time. Furthermore, in the infinite far distance, this vector field should disappear since the size of the source is finite. In this way, the symmetry and the boundary condition provide us with a fundamental constraint on the possible expressions for the vector field and allow us to conduct a general analysis of resulting phenomena. 
Based on this analysis approach, we point out these types of vector-field couplings can change the way the system depends on initial conditions, and also affect the black hole shadow at different wavelengths by a different amount. Accordingly, observations of the shadow in different wavelengths can give rise to the constraints on such couplings. 

The paper is organized as follows. In Sec. \ref{sec:model}, we put forward a phenomenological model which can quantitatively depict how the photon motion is affected by the additional background vector field.
Guided by the symmetry of the Kerr-like spacetime and the boundary condition, in Sec. \ref{sec:photonmotion}, we conduct a general discussion of the motion of photons and in Sec. \ref{sec:shadow} report our general results on the effects of the vector field. In Sec. \ref{sec:results}, we show observational constraints  on model parameters from the EHT experiment by choosing a specific expression for the vector field and constructing the silhouettes of the corresponding black hole. We summarize the main results with a discussion and present a future outlook in Sec. \ref{sec:concl}. We work in natural units where gravitational constant $G=1$ and speed of light $c=1$ and we adopt the metric convention $(-,+,+,+)$.

\section{The model and the stationary axially symmetric rotating spacetime}
\label{sec:model}

We assume the presence of a background vector field $T_{\mu}(X)$, in addition to the spacetime metric tensor field $g_{\mu\nu}(X)$.  This field couples to the photon field and hence deflects null paths. We also assume that the expression of $T_{\mu}(X)$ and the coupling form with the motion of photons cannot be equivalently absorbed into metric $g_{\mu\nu}(X)$ or, in other words, the Levi-Civta connection. So for a free fall, no rotation observer where the Levi-Civta connection equals zero at a given spacetime point, the effects of vector field $T_{\mu}(X)$ would still be present. As a result, this fact could be regraded as that the Einstein's principle of equivalence is violated by the nonmetricity of spacetime. However, we emphasized that if the physical meaning of $T_{\mu}(X)$ does not represent the modification of gravity theories, the violation of Einstein's equivalence principle would just be phenomenological.

The general action describing the coupling between $T_{\mu}(X)$ and the motion of photons is expressed as
\begin{align}
\label{main action}
 S = \int d\lambda \Big[ -\frac{1}{2} e(\lambda)^{-1} g_{\mu\nu} \dot{X}^{\mu} \dot{X}^{\nu} + C\big(T_{\mu}, \dot{X}^{\mu}\big)\Big] ~,
\end{align}
where the dot represents the derivative with respect to the affine parameter $\lambda$ with the mass dimension. The first term is the kinetic term of a test photon in a curved geometry. $e(\lambda)$ is an auxiliary field with the mass dimension, which may help us to eliminate the dimension singularity caused by the massless particle. This idea is similar to the Polyakov action in the string theory \cite{string}. Varying this action with respect to $e(\lambda)$ leads to the constraint equation:
\begin{align}
\label{constrain}
 g_{\mu\nu}\dot{X}^{\mu}\dot{X}^{\nu}=0.
\end{align}
Note that, the auxiliary field $e(\lambda)$ could be absorbed by a redefinition of $\lambda$, making the coefficient in front of the first term constant. This fact will not have any influence on the dynamics of the system so for convenience we eliminate it without affecting the physical results of the present study. $C(T_{\mu},\dot{X}_{\mu})$ depicts that the background vector field $T_{\mu}(X)$ can have coupling with the motion of photons. The coupling leads to an additional force on the photons and can be written as
\begin{align}
 C\big(T_{\mu},\dot{X}^{\mu}\big) = \sum^{\infty}_{n=1} q_n \big(T_{\mu}\dot{X}^{\mu}\big)^n ~,
 \label{general coupling term}
\end{align}
where $q_n$ are dimensional coupling constants. Assuming that the effects of $T_{\mu}$ should be suppressed by the increase of $n$ for a physically feasible form of $T_{\mu}$, we only consider the first power term, which serves our purpose to reveal the phenomenon when $T_\mu$ appears, i.e.,
\begin{align}
 C\big(T_{\mu},\dot{X}^{\mu}\big)=q_1T_{\mu}\dot{X}^{\mu} ~.
 \label{coupling term}
\end{align} 
The dynamics of the background fields $g_{\mu\nu}(X)$ and $T_{\mu}(X)$, are described by the following formal action:
\begin{align}
 S_m =& \int d^4X \sqrt{-g} \Big[ \frac{1}{16\pi} R(g_{\mu\nu}) + D(T_{\mu}) \nonumber \\
 &+ (\textrm{source terms of } g_{\mu\nu} \textrm{ and } T_{\mu} ) \Big] ~,
 \label{sm}
\end{align}
where $R$ is the Ricci scalar, $D$ represents the dynamics of $T_{\mu}$ itself, and we assume no other direct coupling between $g_{\mu\nu}$ and $T_{\mu}$. This paradigm can be analogous to the Kerr-Newman black hole solution of the Einstein-Maxwell theory, where the minimal coupling between $g_{\mu\nu}$ and electromagnetic field introduces only a constant (electric charge) in the metric.

We are interested in the most generic phenomena caused by the vector field $T_{\mu}$, so we shall not adopt a specific form for $D$. Instead, with the assumption that the vector field $T_{\mu}$ effectively has the same excitation source as the curved space-time, $T_{\mu}$ should reflect the same symmetry as the space-time. 
Combining 
this with the known boundary conditions, reasonable expressions for $T_{\mu}$ can be constrained. Let us consider the vacuum situations, since the auxiliary field $e(\lambda)$ could be chosen as a constant, the action Eq. \eqref{main action} with the coupling term Eq. \eqref{coupling term} does not have obvious dependence on the affine parameter $\lambda$, hence the corresponding Hamiltonian is conserved as follows,
\begin{align}
\label{Hamiltonian}
 H &= P_{\mu} \dot{X}^{\mu} - \mathcal{L} = -\frac{1}{2} g_{\mu\nu} \dot{X}^{\mu} \dot{X}^{\nu} \\
 &= -\frac{1}{2} g^{\mu\nu} P_{\mu} P_{\nu} + q_1 g^{\mu\nu} P_{\mu} T_{\nu} - \frac{q_1^2}{2} g^{\mu\nu} T_{\mu} T_{\nu} = \varepsilon ~, \nonumber
\end{align}
where $\varepsilon$ is a constant and $P_{\mu}=\partial\mathcal{L}/\partial\dot{X}^{\mu}$ is the conjugate momentum of the coordinates $X^{\mu}$. This corresponds to the constraint Eq. \eqref{constrain} with $\varepsilon=0$.

We consider a Kerr-like black hole space-time with the line element in the Boyer-Lindquist coordinates $(t,r,\theta,\phi)$ given by
\begin{align}
 ds^2 =& -\Big(1-\frac{2m(r)r}{\rho^2}\Big)dt^2+\frac{\rho^2}{\Delta}dr^2+\rho^2d\theta^2 \nonumber \\
 &+\Big(r^2+a^2+\frac{2a^2m(r)r\sin^2\theta}{\rho^2}\Big)\sin^2\theta d\phi^2 \nonumber \\
 &-\frac{4m(r)ar\sin^2\theta}{\rho^2}d\phi dt ~,
 \label{metric}
\end{align}
where $\Delta(r)\equiv r^{2}-2 m(r) r+a^{2}$, $\rho^{2}\equiv r^{2}+a^{2}\cos^{2}\theta$ and $a$ is the angular momentum per unit mass of black hole $a=J/M$. Note that we have added an additional $r$ dependence in $m(r)$ compared to the Kerr black hole solution in GR, where $m(r)$ is equal to the black hole mass $M$. To recover Minkowski spacetime in the infinite asymptotic distance, $m(r)$ should satisfy
\begin{align}
 \lim_{r \to \infty}\frac{m(r)}{r}=0 ~.
 \label{am}
\end{align}
The function $m(r)$ could describe the modification induced by the minimal coupling between $T_{\mu}$ and $g_{\mu\nu}$ Eq. \eqref{sm}. And out of pure phenomenological interest, $m(r)$ is also related to many kinds of beyond Kerr black holes. For example, when $m(r)=M-C^2/(2r)$, where $C$ is a constant, Eq. (\ref{metric}) describes a Kerr-Newman black hole. The rotating Hayward black hole \cite{Hayward:2005gi, Bambi:2013ufa} and rotating Bardeen black hole \cite{Bambi:2013ufa, Borde:1996df} also have a specific form of $m(r)$, which could avoid the singularity of the ordinary Kerr black hole. Furthermore, the phenomenon brought by the deviation of the Newtonian gravitational constant $G$ could be equivalently described by the function $m(r)$ \cite{Kumar:2019ohr, Held:2019xde}. So keeping the function $m(r)$ is phenomenologically necessary for the current study to compare the effects brought by the vector field $T_{\mu}(X)$ and those brought by the metric with different $m(r)$.

In Boyer-Lindquist coordinate, the time-like and space-like Killing vectors ($\xi_E = \partial_t$ and $\xi_L = \partial_\phi$) describing a stationary axial symmetry spacetime are
\begin{align}
 \xi^{\mu}_E=(1,\ 0,\ 0,\ 0) ~,~ \xi^{\mu}_L=(0,\ 0,\ 0,\ 1) ~,
 \label{Killing vectors}
\end{align}
as the components of the metric field only depend on the coordinates $r$ and $\theta$. Given that the vector field $T_{\mu}$ is  assumed to have the same source as the metric, it is reasonable to think that it should have the same symmetries as the space-time either. So $T_{\mu}$ does not have dependence on the coordinates $t$ and $\phi$. Apart from the Hamiltonian Eq. \eqref{Hamiltonian}, there are two more conservation quantities corresponding to these two Killing vectors Eqs. \eqref{Killing vectors}:
\begin{align}
 E =& \Big( 1-\frac{2m(r)r}{\rho^2} \Big) \dot{t} + \frac{2am(r) r \sin^2\theta}{\rho^2}\dot{\phi} + q_1T_t(r,\theta) ~, 
 \label{E} \\
 L_z =& -\Big( r^2 +a^2 +\frac{2a^2m(r) r \sin^2 \theta}{\rho^2} \Big) \sin^2\theta \dot{\phi} \nonumber \\
 & + \frac{2am(r)r\sin^2\theta}{\rho^2}\dot{t} +q_1T_{\phi}(r,\theta) ~.
\end{align}
These two conservation quantities could be directly derived from the fact that the Lagrangian does not depend on $t$ and $\phi$.

In addition to the continuous symmetries, there is a discrete symmetry which reflects the black hole (as the source of the fields) is a ``rotating body." Since the coordinate system has been fixed to the Boyer-Lindquist form, this symmetry is expressed by the invariance of the metric under the inversion of both $t$ and $\phi$ \cite{Chandrasekhar:1985kt}, which is why the components $g_{tr}$, $g_{t\theta}$, $g_{\theta\phi}$, and $g_{r\phi}$ vanish. Then the assumption that $T_{\mu}$ has the same symmetries as the metric implies $T_{\mu}$ either stays the same or changes signs. So the continuous symmetry Eqs. \eqref{Killing vectors} and the discrete symmetry of the rotating body tell us two possible expressions for the vector field $T_{\mu}$ exist:
\begin{align}
 &\textrm{case I : } \big( T_t(r,\theta),0,0,T_{\phi}(r,\theta) \big) ~,
 \label{caseI} \\
 &\textrm{case II: } \big( 0,T_r(r, \theta),T_{\theta}(r,\theta),0 \big) ~,
 \label{caseII}
\end{align}
where case I changes signs under the discrete symmetry transformation while case II is invariant. Given that we do not provide a specific dynamic for the field $T_{\mu}$ and wish to conduct a general discussion, we cannot determine which case is more reasonable and need to discuss each case separately.

Furthermore, since the size of the black hole source of $T_{\mu}$ is finite, the physically reasonable expression of $T_{\mu}$ must satisfy
\begin{align}
 \lim_{r \to \infty} T_{\mu}(r, \theta) = 0 ~.
 \label{aT}
\end{align}

\section{The motion of photons}
\label{sec:photonmotion}

In this section, we investigate the motion of a photon in the above Kerr-like space-time with the additional background vector field $T_{\mu}$. Using the Hamilton-Jacobi formulation, we can easily find the equations of motion for the photons. First, we introduce the Hamiltonian principal function $S(\lambda,X^{\mu})$ and let
\begin{align}
 P_{\mu}=\frac{\partial S}{\partial X^{\mu}} ~.
 \label{P}
\end{align}
The Hamilton-Jacobi equation is
\begin{align}
 H + \frac{\partial S}{\partial \lambda} = 0 ~.
 \label{hj}
\end{align}
In the following, we will discuss the above two cases Eps. \eqref{caseI} and \eqref{caseII} individually. 

\subsection{Case I}

Substituting the Hamiltonian Ep. \eqref{Hamiltonian}, the metric Ep. \eqref{metric} and the conjugate momentum Ep. \eqref{P} into the Hamilton-Jacobi Ep. \eqref{hj}, one gets
\begin{align}
\label{HJ}
 -2 \rho^2 \frac{\partial S}{\partial \lambda} &= \frac{1}{\Delta} \big[ (r^2+a^2) \frac{\partial S}{\partial t} +a \frac{\partial S}{\partial \phi} \big]^2 \\
 & -\frac{1}{\sin^2\theta} \big( a \sin^2\theta \frac{\partial S}{\partial t}+\frac{\partial S}{\partial \phi} \big)^2 -\Delta \big( \frac{\partial S}{\partial r} \big)^2 - \big( \frac{\partial S}{\partial \theta} \big)^2 \nonumber \\
 &+2q_1g_E(r,\theta)\frac{\partial S}{\partial t}+2q_1g_L(r,\theta)\frac{\partial S}{\partial\phi}+q_1^2g_T(r,\theta) ~, \nonumber
\end{align}
where we have defined
\begin{align}
 g_E(r, \theta) &\equiv \frac{-\Sigma^2}{\Delta} T_t(r, \theta) -\frac{2 a m(r) r}{\Delta} T_{\phi}(r, \theta) ~, \label{ge} \\
 g_L(r, \theta) &\equiv \frac{\Delta - a^2 \sin^2\theta}{\Delta \sin^2\theta} T_{\phi}(r, \theta) -\frac{2 a m(r) r}{\Delta} T_t(r, \theta) ~, \label{gl} \\
 g_T(r, \theta) &\equiv \frac{\Sigma^2}{\Delta} T_t(r, \theta)^2 - \frac{\Delta -a^2 \sin^2\theta}{\Delta \sin^2\theta} T_{\phi}^2 \nonumber \\
 & + 4 \frac{a m(r) r}{\Delta} T_t(r, \theta) T_{\phi}(r, \theta) ~, \label{gt}
\end{align}
and
\begin{align}
 \Sigma^2 \equiv (r^2+a^2)^2-\Delta a^2\sin^2\theta ~. \nonumber
\end{align}
For the Kerr space-time in GR, apart from $\varepsilon$, $E$, $L_z$, there is another conservation quantity: the 
Carter 
constant $\mathcal{K}$. When the additional vector field $T_{\mu}$ is present, this point cannot be guaranteed, making things difficult to solve analytically. However, if we consider the situation that $g_E$, $g_L$, and $g_T$ can be decomposed as
\begin{align}
 g_E(r,\theta) &= g_E^r(r)+g_E^\theta(\theta) ~, \nonumber \\
 g_L(r,\theta) &= g_L^r(r)+g_L^\theta(\theta) ~, \nonumber \\
 g_T(r,\theta) &= g_T^r(r)+g_T^\theta(\theta) ~, \label{decomposition}
\end{align}
the fourth integral constant will still appear. According to the asymptotic behaviors Eps. \eqref{am}, \eqref{aT} and the definitions Eps. \eqref{ge}, \eqref{gl}, \eqref{gt}, we obtain $g_L^{\theta}=0$ and the asymptotic behaviours
\begin{align}
 \lim_{r \to \infty} \frac{g_E^r(r)}{r^2} = \lim_{r \to \infty} g_L^r(r) = \lim_{r \to \infty} \frac{g_T^r(r)}{r^2} = 0 ~.
\label{ag}
\end{align}
We now assume that the Hamilton principal function $S$ has the following form:
\begin{align}
 S(\lambda,t,r,\theta,\phi)=\frac{\varepsilon \lambda}{2} -E t+L_{z} \phi+S_{r}(r)+S_{\theta}(\theta) ~.
\label{S}
\end{align}
Substituting this expression into Eq. \eqref{HJ}, one obtains
\begin{align}
	&\Delta\left(\frac{dS_r}{dr}\right)^2-\frac{1}{\Delta}\left[(r^2+a^2)E-aL_z\right]^2+(L_z-aE)^2-\varepsilon r^2 \nonumber \\
	&+2q_1\left[g_E^r(r)E-g_L^r(r)L_z\right]-q_1^2g_T^r(r) \nonumber \\
	&=-\left(\frac{dS_{\theta}}{d\theta}\right)^2-(L_z^2\sin^{-2}\theta-a^2E^2)\cos^2\theta+\varepsilon a^2\cos^2\theta \nonumber \\
	&-2q_1g_E^\theta(\theta)E-q_1^2g_T^\theta(\theta) ~.
	\label{Seq}
\end{align}
The left-hand side of the equation only depends on the coordinate $r$ and the right-hand side of the equation only depends on the coordinate $\theta$, which implies these two parts must be equal to the same constant $\mathcal{K}$. So in addition to $\varepsilon$, $E$, and $L_z$, we obtain the fourth integral constant $\mathcal{K}$, which shows the assumed $S$ Eq. \eqref{S} is self-consistent according to the Hamilton-Jacobi formulation. By integrating the left side and right side of Eq. \eqref{Seq} respectively, we obtain expressions for $R(r)$, $\Theta(\theta)$ and the Hamiltonian principal function $S$,
\begin{align}
 S =& \frac{1}{2}\varepsilon\lambda-Et +L_z\phi \nonumber \\ 
 &+\sigma_r\int^r\frac{\sqrt{R(r)}}{\Delta}dr+\sigma_\theta\int^\theta\sqrt{\Theta(\theta)}d\theta ~,
\end{align}
where $\sigma_r=\pm1$, $\sigma_\theta=\pm1$. The definitions of $R(r)$, $\Theta(\theta)$ are
\begin{align}
 R(r) =& E^2(r^2+a^2-a\xi)^2 -\Delta E^2\Big[\eta+(\xi-a)^2-\frac{\varepsilon r^2}{E^2}\nonumber \\
 & +2\frac{q_1}{E}g_E^r(r)-2\frac{q_1}{E}\xi g_L^r(r)-\frac{q_1^2}{E^2}g_T^r(r)\Big] ~, \label{Rr} \\
 \Theta(\theta) =& E^2 \Big[ \eta -\big( \xi^2 \sin^{-2}\theta -a^2 -\frac{\varepsilon a^2}{E^2} \big) \cos^2\theta -2 \frac{q_1}{E}g_E^\theta(\theta) \nonumber \\
 & +\frac{q_1^2}{E^2} g_T^\theta(\theta) \Big] ~, \label{Thetatheta}
\end{align}
where we have defined
\begin{align}
	\xi \equiv\frac{L_z}{E} ~,~ \eta \equiv\frac{\mathcal{K}}{E^2} ~.
\end{align}
We have thus obtained the Hamiltonian principal function expressed as the function of the coordinates $(t,\ r,\ \theta,\ \phi)$ and the integral constants $\varepsilon$, $E$, $L_z$, $\mathcal{K}$. The equations of motion are completely determined:
\begin{align}
 \rho^2\dot{r} =& -\frac{1}{\sigma_r}\sqrt{R(r)} ~, \label{r} \\
 \rho^2\dot{\theta} =& -\frac{1}{\sigma_\theta}\sqrt{\Theta(\theta)} ~, \label{theta} \\
 \rho^2\dot{\phi} =& -\frac{E}{\Delta}\left[2am(r)r+(\rho^2-2m(r)r)\xi\sin^{-2}\theta\right] \nonumber \\
 & +q_1g_L^r(r) ~, \label{phi} \\
 \rho^2\dot{t} =& -\frac{E}{\Delta}\left(\Sigma^2-2am(r)r\xi\right)+q_1g_E^r(r)+q_1g_E^\theta(\theta) ~.
\end{align}
Note that for the motion of photons $\varepsilon=0$.

\subsection{Case II}

Similar to case I, for the situation that $T_r$, $T_\theta$ satisfy
\begin{align}
 T_r(r,\theta) = T_r(r) ~,~ T_\theta(r,\theta) = T_\theta(\theta) ~,
\label{decompositionII}
\end{align}
the Hamiltonian principal function $S$ of case II also has a separable solution Eq. \eqref{S}. According to Eq. \eqref{aT}, $T_{\theta}(\theta)=0$. The corresponding Hamilton-Jacobi equation takes
\begin{align}
 &\Delta \Big( \frac{dS_r}{dr} -qT_r(r) \Big)^2 -\frac{1}{\Delta} \big[ (r^2+a^2)E -aL_z \big]^2 \nonumber \\
 &+(L_z-aE)^2 -\varepsilon r^2 \\
 &= -\big( \frac{dS_\theta}{d\theta} \big)^2 -(L_z^2\sin^{-2}\theta -a^2E^2) \cos^2\theta +\varepsilon a^2 \cos^2\theta ~, \nonumber
\end{align}
which leads to the following expression of $S$:
\begin{align}
 S &= \frac{1}{2} \varepsilon \lambda - Et + L_z\phi + \int^r \Big[ \sigma_r\frac{\sqrt{R(r)}}{\Delta}+q_1T_r(r) \Big] dr \nonumber \\
 &+ \int^\theta \sigma_\theta \sqrt{\Theta(\theta)} d\theta ~,
\label{SII}
\end{align}
where
$$
 R(r) = E^2(r^2+a^2-a\xi)^2 -\Delta E^2 \Big[ \eta+ \big( \xi-a \big)^2-\frac{\varepsilon r^2}{E^2} \Big] ~,
$$
and 
$$
 \Theta(\theta) = E^2 \Big[ \eta-\Big( \xi^2\sin^{-2}\theta-a^2-\frac{\varepsilon a^2}{E^2} \Big) \cos^2\theta \Big] ~.
$$
So we could find that the only modification to the Hamiltonian principal function is the term $q_1T_r(r)$, which is present in the Hamiltonian principal function Eq. \eqref{SII}. According to the Hamiltonian-Jacobi formulation, since this term does not involve any integral constants, it will not change the equations of motion. So for the case II, with $T_{\mu}$ having the form Eqs. \eqref{decompositionII}, the additional vector field does not have influence on the motion of photons, i.e.,
\begin{align}
 &\rho^2\dot{r}=-\frac{1}{\sigma_r}\sqrt{R(r)} ~, \label{rII} \\
 &\rho^2\dot{\theta}=-\frac{1}{\sigma_\theta}\sqrt{\Theta(\theta)} ~, \label{thetaII} \\
 &\rho^2\dot{\phi}=-\frac{E}{\Delta}\left[2am(r)r+(\rho^2-2m(r)r)\xi\sin^{-2}\theta\right] ~, \label{phiII} \\
 &\rho^2\dot{t}=-\frac{E}{\Delta}\left(\Sigma^2-2am(r)r\xi\right) ~.
\end{align}
This result is based on the assumption Eqs. \eqref{decompositionII}. For a general form of $T_r(r,\theta)$, $T_{\theta}(r,\theta)$, the absence of $T_{\mu}$'s effects cannot be guaranteed.

\section{The shadow of black holes}
\label{sec:shadow}

Now we discuss how to determine the apparent shape of the rotating black hole shadow. Let us consider an observer at a large distance from the black hole along an inclination angle $\theta_0$ between the rotation axis of the black hole and the line of sight of the distant observer. The contour of the shadow can be expressed by celestial coordinates $\alpha$ and $\beta$, where $\alpha$ corresponds to the apparent perpendicular distance of the shape as seen from the axis of the symmetry, and $\beta$ is the apparent perpendicular distance of the shape from its projection on the equatorial plane. The expressions of $\alpha$ and $\beta$ can be determined from the geometry \cite{Chandrasekhar:1985kt}:
\begin{align}
 \alpha &= \lim_{r \to \infty} \big( -r^2\sin\theta\frac{d\phi}{dr} \big) \Big\vert_{\theta=\theta_0} ~, \nonumber \\
 \beta &= \lim_{r \to \infty} r^2 \frac{d\theta}{dr} \Big\vert_{\theta=\theta_0} ~.
\end{align}
Combining the above expressions with Eqs. \eqref{r}, \eqref{theta}, \eqref{phi}, \eqref{rII}, \eqref{thetaII}, \eqref{phiII}, and the asymptotical behaviors Eqs. \eqref{am}, \eqref{ag}, one can derive the expressions for both case I and case II:
\begin{align}
 \alpha &= -\frac{1}{\sin\theta_0} \xi ~, \nonumber \\
 \beta &= \pm \Big[ \eta +(a-\xi)^2 -\big( a\sin\theta_0 -\frac{\xi}{\sin\theta_0} \big)^2 \Big]^{\frac{1}{2}} ~.
\label{shadow}
\end{align}
The shape of the shadow is determined by the unstable orbits with constant radius since they are the boundaries that separate the unbound and bound orbits, which must satisfy
\begin{align}
 R(r) |_{r=r_0} = R'(r) |_{r=r_0} = 0 ~,
\end{align}
where $r_0$ is the radius of the unstable orbits and prime denotes the derivative with respect to $r$. For case I, these two conditions yield
\begin{align}
\label{xieta1}
 &(r_0^2+a^2-a\xi)^2-\Delta(r_0)\Big[\eta+(\xi-a)^2+2\frac{q_1}{E}g^r_E(r_0) \nonumber \\
 &-2\frac{q_1}{E}\xi g^r_L(r_0)-\frac{q_1^2}{E^2}g^r_T(r_0)\Big]=0 ~, 
\end{align} 
and
\begin{align}
\label{xieta2}
 &4r_0(r_0^2+a^2-a\xi)-2(r_0-m'(r_0)r_0-m(r_0)) \Big[ \eta +(\xi-a)^2 \nonumber \\
 &+2\frac{q_1}{E}g^r_E(r_0) -2\frac{q_1}{E}\xi g^r_L(r_0) -\frac{q_1^2}{E^2}g^r_T(r_0) \Big] \\
 &-\Delta(r_0) \Big[ 2\frac{q_1}{E}g_E^r{'}(r_0) -2\frac{q_1}{E}\xi g^r_L{'}(r_0) -\frac{q_1^2}{E^2}g^r_T{'}(r_0) \Big] =0 ~. \nonumber
\end{align}
Solving these two equations and ignoring non-physical solutions, one gets $\xi$, $\eta$ expressed as functions of $r_0$, $E$, i.e., $\xi(r_0,E)$, $\eta(r_0,E)$. The corresponding celestial coordinates $\alpha$ and $\beta$ can be derived using Eqs. \eqref{shadow}. 

Note that, the dependence of $\xi$, $\eta$ on the integral constant $E$ is the direct result of introducing the additional vector field $T_{\mu}$. When $T_{\mu}$ is absent, all the $g(r)$ functions in Eqs. \eqref{xieta1} and \eqref{xieta2} vanish and the solutions of $\xi$, $\eta$ will only have $r_0$ dependence, i.e., $\xi(r_0)$, $\eta({r_0})$. According to the expression Eq. \eqref{E}, for the distant observer, the integral constant $E$ is the photon's energy. This fact tells us that different frequencies of light will give rise to different features in the black hole shadow. So it is feasible to test the existence of the vector field $T_{\mu}$ by observing the shadow of a black hole in multiple wavelengths.

Finally, we want to point out that the above conclusion is based on the assumption that the Hamiltonian principal function has a separable solution Eq. \eqref{S}. This assumption guarantees the existence of unstable photon orbits with constant radius and the induced black hole shadow, which is thus consistent with the observation of a dark area surrounded by a bright emission ring. In the situation that Eq. \eqref{S} cannot be separated, there might also have unstable photon orbits and the formation of the black shadow according to the analysis of space-time separability given in Refs. \cite{Glampedakis:2018blj, Shaikh:2019fpu}; however, we should note that the physical reason for the presence of the dependence on energy $E$ is that the vector field $T_{\mu}$ with the first power coupling term Eq. \eqref{coupling term} introduces the first power of the velocity $\dot{X}_{\mu}$, which thus changes the way that the system depends on the integral constants or, in other words, the initial conditions. This can be more easily seen by Eq. \eqref{Seq}, where the terms containing the integral constants $E$, $L_z$ to the first power are introduced. Therefore, even for the situation that the vector field $T_{\mu}$ has a non-separable Hamiltonian principal function $S$, the result of a frequency-dependent shadow may remain, assuming the black hole shadow still exists. Furthermore, for a more general coupling term Eq. \eqref{general coupling term}, one might expect that the odd power terms would introduce the energy dependence while the even power terms will not.

\section{An example and the observational constraints}
\label{sec:results}

In this section, we choose a particular expression for $T_{\mu}$  to illustrate the specific effects brought by the vector field $T_{\mu}$ and the relevant observational constraints.

Inspired by the movement of a charged particle around the Kerr-Newman black hole \cite{Vladimir19893, Vladimir19896, Hackmann:2013pva}, we know an expression for $T_{\mu}$ that makes the Hamiltonian have a separable solution Eq. \eqref{S},
\begin{align}
\label{vector}
 T_{\mu}=\frac{Q f(r)}{\rho^2}(-1,\ 0,\ 0,\ a\sin^2\theta) ~,
\end{align}
where $Q$ is a constant describing the quality of the source, $f(r)$ is a function determined by the underlying fundamental theory, whose expression should satisfy the boundary condition Eq. \eqref{aT}. Then, according to Eqs. \eqref{ge}, \eqref{gl}, \eqref{gt}, \eqref{decomposition}, we have
\begin{align}
 g_E^r(r) &= \frac{Q f(r)(r^2+a^2)}{\Delta} ~,\nonumber \\
 g_L^r(r) &= \frac{Q f(r)a}{\Delta} ~, \nonumber \\
 g_T^r(r) &= \frac{Q^2 f^2(r)}{\Delta} ~,
\end{align}
and $g_E^{\theta}(\theta)=g_L^{\theta}(\theta)=g_T^{\theta}(\theta)=0$.  The conditions for the unstable spherical orbits in Eqs. \eqref{xieta1} and \eqref{xieta2} give rise to
\begin{align}
\label{xieta}
 &\big[ r_0^2+a^2-a\xi-q_E f(r_0) \big]^2 -\Delta(r_0) \big[ \eta+(\xi-a)^2 \big] =0 ~, \nonumber \\
 &2 \big[ r_0^2+a^2-a\xi-q_E f(r_0) \big] \big[ 2r_0-q_Ef'(r_0) \big] \nonumber \\
 &- \Delta'(r_0) \big[ \eta+(\xi-a)^2 \big] = 0 ~,
\end{align}
where $\Delta'(r_0)$ represents the derivative with respect to $r_0$ and we have defined
\begin{align}
\label{qe}
 q_E\equiv q_1\frac{Q}{E} ~.
\end{align} 

\begin{figure*}
\subfigure[{Fixed spin $a/M = 0.1$}]{
	\label{Fig1.sub.1}
	\includegraphics[width=.31\textwidth]{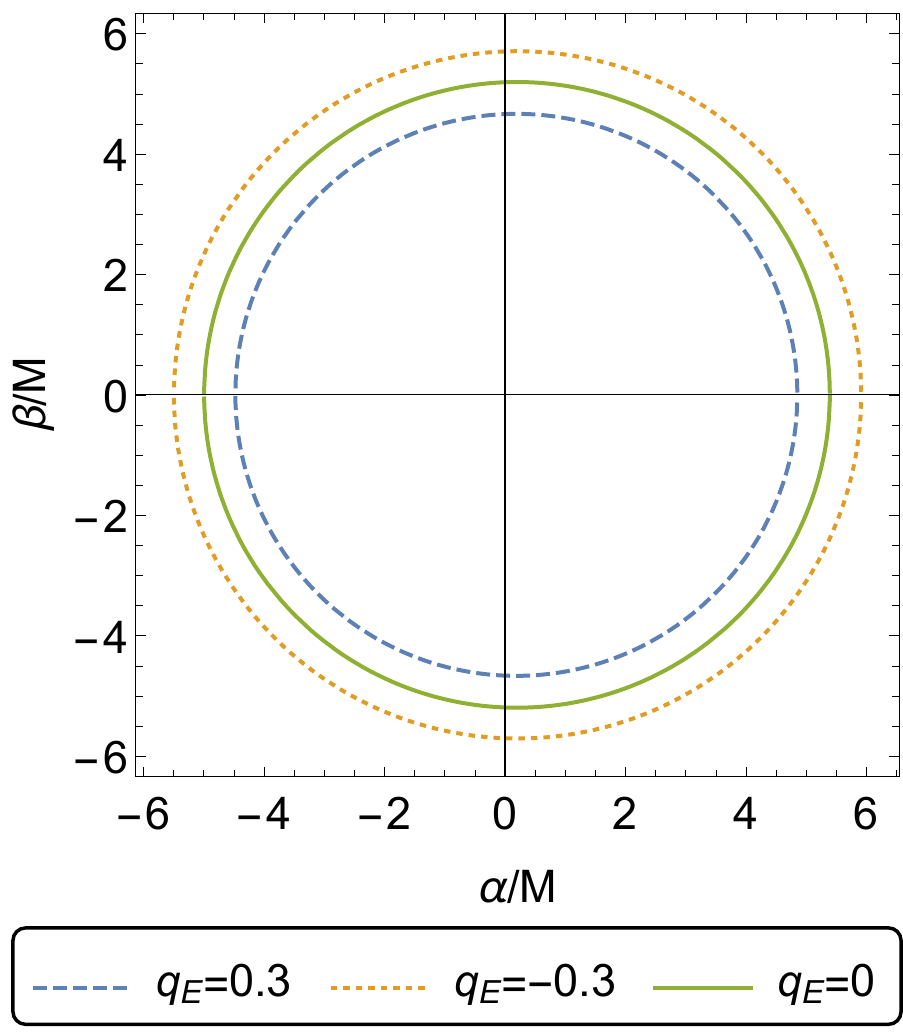}}
\subfigure[{Fixed spin $a/M = 0.7$}]{
	\label{Fig1.sub.2}
	\includegraphics[width=.31\textwidth]{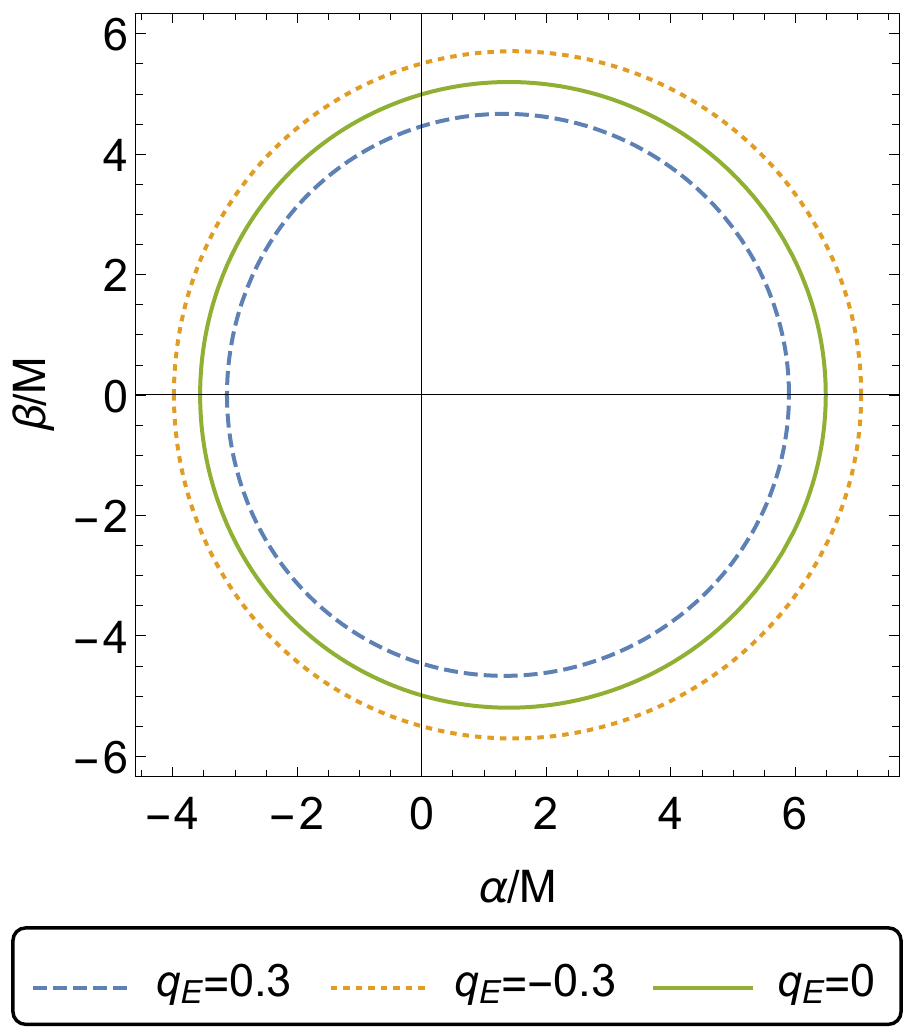}}
\subfigure[{Fixed spin $a/M = 0.99$}]{
	\label{Fig1.sub.3}
	\includegraphics[width=.31\textwidth]{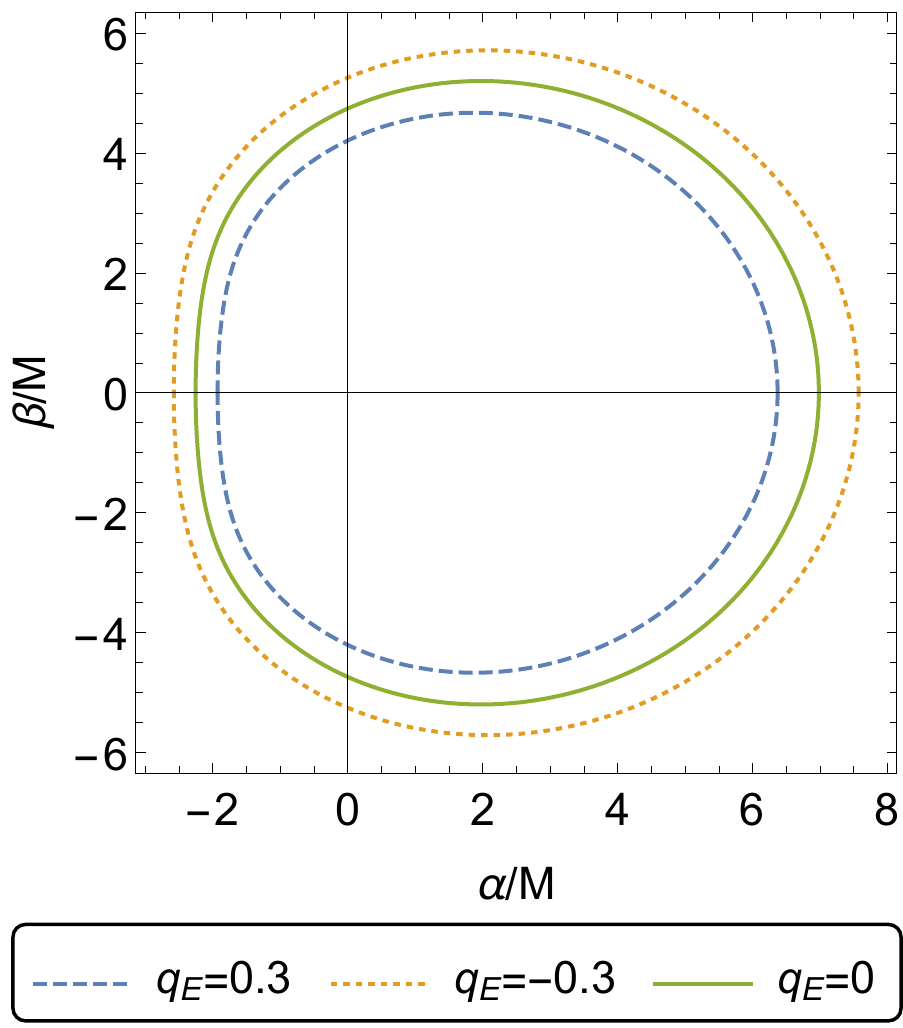}}
\caption{Plot showing the shadow cast by the Kerr black hole $m(r)=M$ with the vector field Eq. \eqref{vector} having the form of $f(r)=r$. The inclination angle $\theta_0$ has been set to $\pi/2$.}
\label{Kerr}
\end{figure*}

\begin{figure}
\includegraphics[width=.45\textwidth]{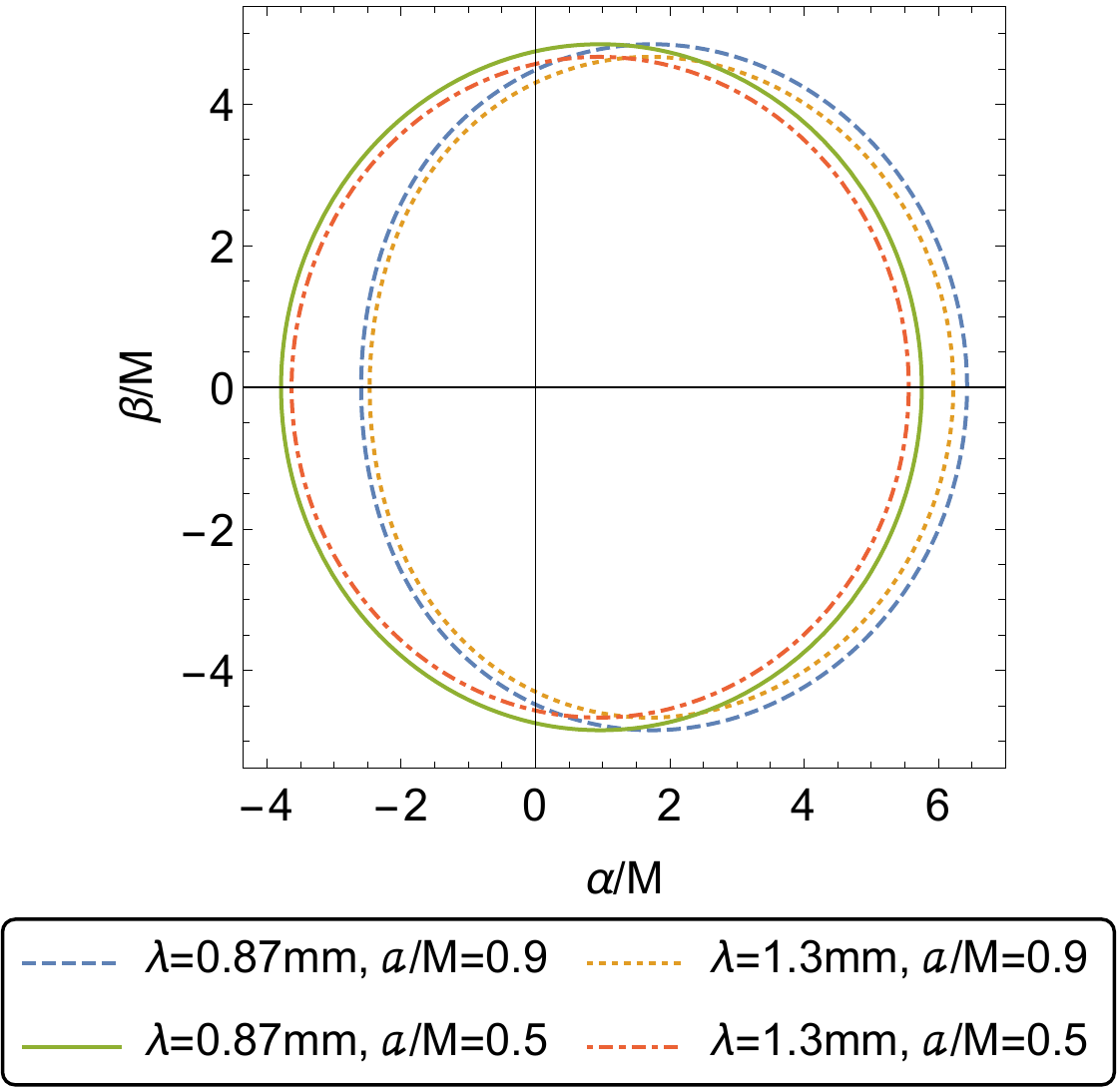}
\caption{This figure has the same settings as Fig.~\ref{Kerr}. Here we plot the shadow of two wavelengths for two spin parameters.}
\label{TF}
\end{figure}

The parameter $Q$ could contribute to the total energy curving the space-time by the possible coupling between $T_{\mu}$ and $g_{\mu\nu}$ shown in Eq. \eqref{sm} and thus modify the standard Kerr metric. However, in physically plausible situations, this contribution to the space-time curvature must be negligibly small compared to the matter contribution of the matter stress tensor. Therefore, to visualize the shadow cast by the rotating black hole with $T_{\mu}$ from Eq. \eqref{vector}, we consider the standard Kerr black hole $m(r)=M$ and generate plots for the coordinates $\alpha/M$ and $\beta/M$ by assuming the approximate behavior $f(r)=r$ at the related scale. These plots are shown in Fig.~\ref{Kerr} for the fixed values of spin parameter $a/M$ and different values of parameter $q_E/M$. It is easy to see the effects of parameter $q_E$ on the shadow: Comparing with the overall size, the shape of the black hole shadow is slightly changed. An increase in the value of $q_E$ decreases the overall size of the shadow. The specific value of parameter $q_E$ is related to the frequency of the photons that we observed according to the definition Eq. \eqref{qe}, which is consistent with the conclusion of the last section, i.e., the black hole shadow will have a different appearance under different 
wavelengths. 
In Fig.~\ref{TF}, we show this point more clearly. Assuming that $q_E=0.3$ at $\lambda=1.3$mm, we plot the appearances of the black hole shadow corresponding to the 1.3mm (230GHz) and 0.87mm (345GHz) wavelengths, respectively. These two wavelengths are able to be simultaneously observed by the next-generation Event Horizon Telescope (ngEHT) \cite{Blackburn:2019bly}. Thus, the scenario under consideration is promising to be tested by observations in the near future.

\begin{figure}
\includegraphics[width=.42\textwidth]{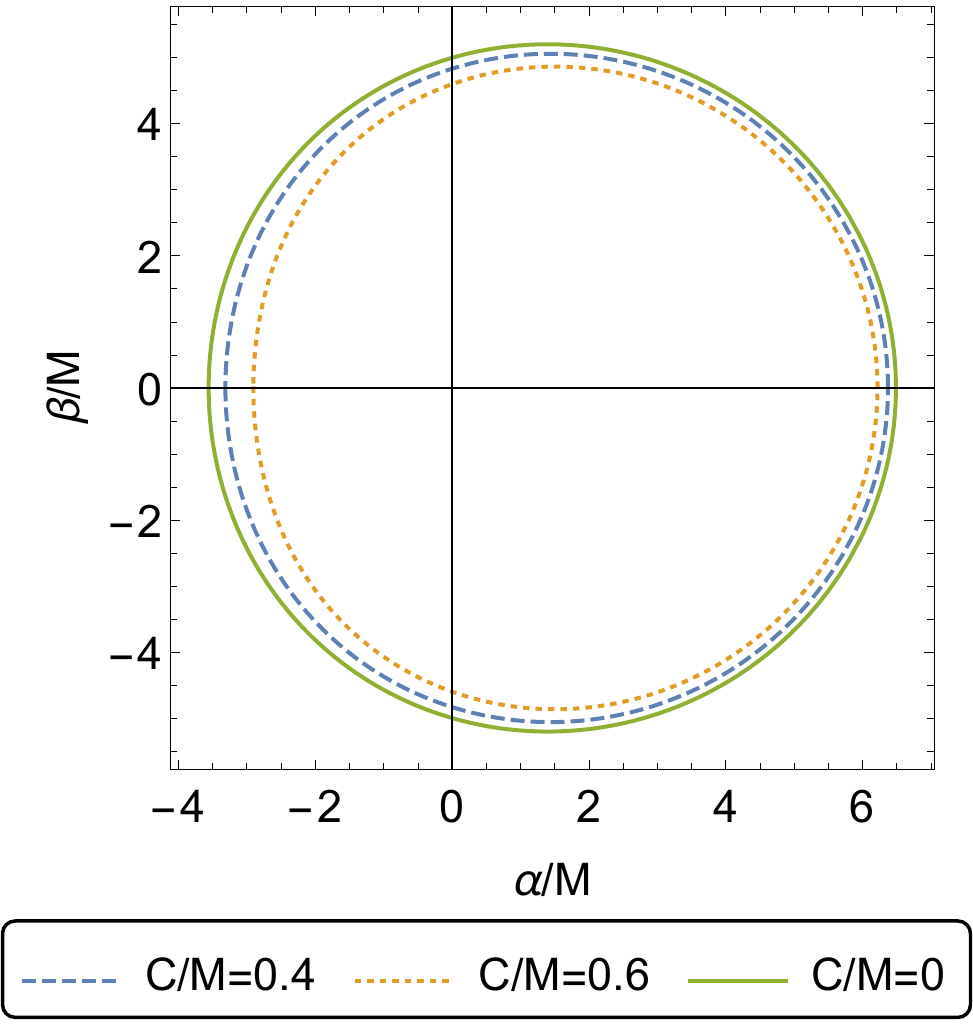}
\caption{Plot showing the shadow cast by the Kerr-Newman black hole in the absence of $T_{\mu}$. 
Here we have fixed spin parameter $a/M = 0.7$.
}
\label{KN}
\end{figure}

\begin{figure}
\includegraphics[width=.43\textwidth]{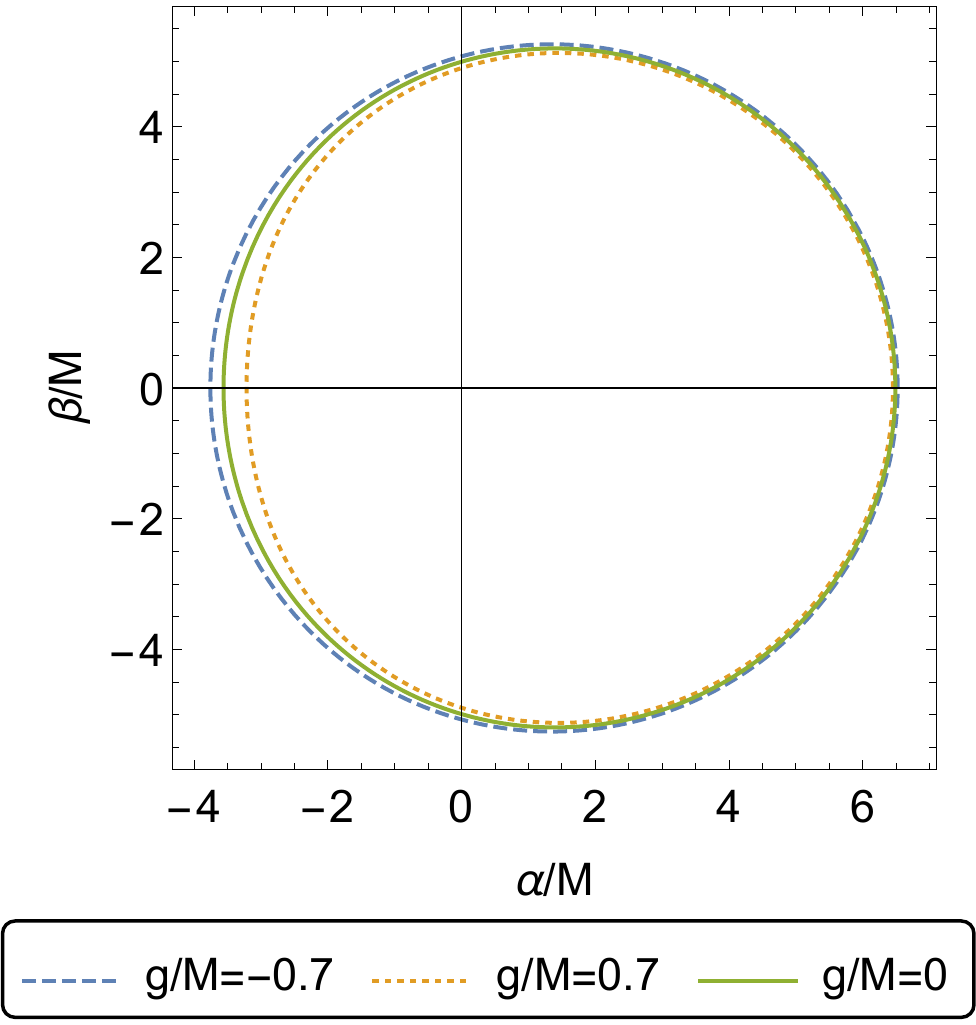}
\caption{Plot showing the shadow cast by the nonsingular rotating Hayward black hole in the absence of $T_{\mu}$. 
Here we have fixed spin parameter $a/M = 0.7$.
}
\label{NH}
\end{figure}

We now make a comparison with the effects caused by the function $m(r)$ since it describes a large class of 
deviations 
from the standard Kerr black hole and has been widely discussed in the literature. In the absence of $T_{\mu}$, i.e., $q_E=0$, let us first consider the Kerr-Newman black hole:
\begin{align}
 m(r) = M -\frac{C^2}{2r} ~,
\end{align}
Figure ~\ref{KN} displays the corresponding shadow for different values of parameter $C$. Another typical expression of $m(r)$ is
\begin{align}
 m(r) = M \frac{r^3}{r^3 +g^3} ~,
\end{align}
where $g$ is the model parameter. This form can describe the nonsingular rotating Hayward black hole \cite{Hayward:2005gi} and a type of asymptotically safe gravity \cite{Held:2019xde}. Figure ~\ref{NH} displays the corresponding shadow for different values of parameter $g$.

According to Figs. ~\ref{KN} and ~\ref{NH}, we note that the obvious deviation from the standard silhouette only occurs on the left-hand side of the picture. This side corresponds to the spherical orbits that have relatively small radius $r_0$. According to Eqs. \eqref{xieta}, the $m(r_0)$ function only appears in the function $\Delta(r_0)$ and its derivative, which means the largest modification by $m(r_0)$ only occurs on orbits with the smallest radius $r_0$ given by the condition Eq. \eqref{am}. Therefore, the fact that the most significant deformation only occurs on one side of the shadow is a general result for the function $m(r)$, which is different from the effects caused by $T_{\mu}$ with the form Eq. \eqref{vector} where only the overall shadow size is adjusted.

In the current study, we have set the inclination angle $\theta_0=\pi/2$ between the rotation axis and the line of sight. It must be difficult to realize this in the real world. Fortunately, a detailed numerical study \cite{Chan:2013gez} based on the geometrical relations such as Eqs. (\ref{shadow}) shows that choosing a different inclination angle mainly changes the shape of the shadow by an overall horizontal displacement, while the overall size, i.e., the average radius of the pattern is almost unchanged. Therefore, it makes sense to focus on the overall size of shadow since the current observations do not yet yield detailed information about the shape characteristics of a black hole shadow.

Finally, let us consider the current constraints on the model parameter $q_E$. 
According to the above results, we have learned that the crucial test of the scenario in Eq. \eqref{main action} is to see whether the black hole shadow has dependence on the observed wavelength. However, the current EHT experiment only operates at a wavelength of $1.3$ mm \cite{Akiyama:2019cqa}. 
Since the effect of appreciable $q_E$ is mainly to change the overall size of the shadow, a probe for this parameter would be measurement of the black hole mass $M$. That is, $q_E$ can be constrained by measurement at different scales on the black hole mass $M$. Specifically,  stellar-dynamics observations could provide a measurement on the mass $M_{stellar}$ of the SMBH in the Newtonian gravity approximation. And the observations of the black hole shadow provide us with another measurement of mass $M_{shadow}$ by the physics at the horizon scale. In principle, the contrast between these two measurements $M_{stellar}$ and $M_{shadow}$ could place constraints on the parameter $q_E$ if assuming $T_{\mu}$ plays no role beyond the horizon scale.

We begin with the Schwarzschild black hole, i.e., $a=0$. According to the spherical symmetry, the unstable null spherical orbits should be confined to a plane, i.e., forming a circular orbit. Then we can choose $\theta=90^{\circ}$, $\dot{\theta}=0$; Eqs. \eqref{xieta} becomes
\begin{align}
 &(r_0^2-q_Er_0)^2-(r_0^2-2Mr_0)\xi^2=0 ~, \nonumber \\
 &(r_0^2-q_Er_0)(2r_0-q_E)-(r_0-M)\xi^2=0 ~.
\label{spherical}
\end{align}
The size of the shadow is given by:
\begin{align}
 d =2|\xi| = \bm{\alpha} \big( M+\Delta M \big) ~,
\end{align}
where we have defined
\begin{align}
 \Delta M = \bm{g}(q_E)M ~,
\end{align}
and $\bm{g}$ is a function with respect to $q_E$ derived from Eqs. \eqref{spherical}. $\bm{\alpha}$ is the parameter describing the difference between the size of the shadow and the gravitational radius $M$, which contains the influence of the inclination $\theta_0$ and the spin parameter $a$ on the size of the shadow, i.e. the deviation from the Schwarzschild black hole. For the current Schwarzschild case, $\bm{\alpha}=6\sqrt{3}$. In the EHT observation for the M87* black hole, $\bm{\alpha}=11^{+0.5}_{-0.3}$ is obtained by fitting the observed shape models to a large number of visibility data generated from the Image Library \cite{Akiyama:2019cqa}. This should lead to a slightly different $\bm{g}$ function from that of $\bm{\alpha}\approx 10.4$. However since $f(r)$ does not introduce any special dependence on the spin parameter $a$, this difference must be negligible. Finally, we obtain the relationship between the mass $M_{stellar}$ measured by the dynamical methods and the mass $M_{shadow}$ measured by the optical shadow:
\begin{align}
 M_{stellar} + \Delta M_{stellar} = M_{shadow} ~.
\end{align}
For the M87* black hole, recent stellar-dynamics observations by Gebhardt et al. \cite{Gebhardt:2011yw, Gebhardt:2009cr} found $M_{stellar} = (6.6 \pm 0.4) \times 10^9M_\odot$ and the EHT experiments derive $M_{shadow} = (6.5\pm 0.7) \times 10^9 M_\odot$ \cite{Akiyama:2019cqa}. This leads to the sub-maximum range for $q_E$:
\begin{align}
 q_E \in (-0.50,\ 0.56) ~.
\end{align}
Furthermore, the gas dynamics observations give a rather different measurement of the mass of the M87* black hole $M_{gas} = 3.5^{+0.9}_{-0.3} \times 10^9 M_\odot$ \cite{Harms:1994eh, Macchetto:1997gi, Walsh:2013uua}, which would lead to a weaker constraint on the parameter $q_E$. 
Therefore, given the unknown systematic error on the mass measurement, we are looking forward to the ngEHT with the dual wavelength observation capability, which could yield better constraints \cite{Blackburn:2019bly}.

\section{Conclusions}
\label{sec:concl}

In this paper, we have proposed a mechanism for testing the equivalence principle by analyzing black hole shadows. In particular, for rotating black holes which are of high astronomical interest, the features imprinted on their shadow under the influence of an additional vector field $T_{\mu}(X)$, which phenomenologically depicts a violation of equivalence principle, can affect the motions of photons. 
Accordingly, our scenario provides an interesting example to discuss this effect in regions of extremely strong gravitational fields. We assume that $T_{\mu}(X)$ is regarded as a background vector field generated by the central black hole so the symmetries possessed by the black hole and the spacetime could be used to constrain the form of this vector field. Furthermore, we demand the boundary condition that the vector field $T_{\mu}(X)$ vanishes at infinity since it is generated by a finite size source. Under these two constraints, we perform a general analysis on the black hole shadow influenced by $T_{\mu}(X)$ with the coupling form Eq. \eqref{coupling term}. Our key result is that the shadow in edge-on view will have different appearances for different frequencies of the observed light. The physical reason for this phenomenon is that the coupling form shown in Eq. \eqref{coupling term} alters the way that the system depends on the initial conditions by introducing the first power of the velocity $\dot{X}$. Therefore, this phenomenon is quite generic and  is not sensitive to a specific form of $T_{\mu}(X)$.

The current EHT experiments operate at a wavelength of $1.3$mm. Although each station of EHT receives two adjacent frequency bands centered at $227.1$GHZ and $229.1$GHz respectively, these two frequency bands are handled by different groups and are used to eliminate the error of doing correlation among the data \cite{Akiyama:2019cqa}. 
Therefore, current experimental conditions might not allow us to directly determine the existence of the phenomenon: that the observed appearance of the black hole shadow could change with the observed 
wavelengths. It deserves mentioning that the future project of ngEHT could have the ability to observe the 1.3-mm and 0.87-mm wavelengths simultaneously \cite{Blackburn:2019bly}, we hope this project together with other future multi-band observations as well as the related data-processing techniques could allow for tests of this phenomenon.

As an example, we chose the vector field $T_{\mu}$ in the form Eq. \eqref{vector} and studied its effects on the shadow cast by the Kerr black hole. The results show that the overall size of the black hole shadow is altered, which is different from the effect brought by the usual modification on the metric using the function $m(r)$, where only one side of the silhouette has obvious distortion. Thus, there is a large degeneracy between the black hole mass and the model parameter $q_E$. Finally, by using the measurements on the black hole mass at different distance scales, we set constraints on the coupling parameter $\left|q_E<0.5\right|$ by combining the results of EHT and orbits of stars or gas.

We emphasize that, in principle, the black hole shadow characterized by the parameters $\xi$, $\eta$ could be totally determined by the fundamental equations governing the motion of photons. The accretion disk surrounding the black hole only serves to provide a light source and thus the astrophysical complications introduced by the specific accretion model may be avoided \cite{Narayan:2019imo}. However, from a practical observational point of view, the accretion flow may obscure the shadow and this problem would be more serious had the accretion flow been optically thick \cite{Luminet:1979nyg, Fukue, Psaltis:2018xkc}. Although Sgr A* and M87* are not this case, the detailed analyses involving future high-precision measurements should take the influence of the accretion disk into consideration. We note that Refs. \cite{Yuan:2009am, Johannsen:2010ru, Johannsen:2015hib, Psaltis:2014mca, Gralla:2019xty} also studied various features of the shadow.

\section*{Acknowledgments}
We are grateful to Z.H. Fan, J.O. Gong, G. Kang, S. Mukohyama, E. Saridakis, M. Sasaki, Z.Q. Shen, Y.Q. Xue, S.X. Tian, and M. Yamaguchi for stimulating discussions and valuable comments.
HZ acknowledges support from the USTC fellowship for international visiting professors and from Shanghai Astronomical Observatory. 
This work is supported in part by the NSFC (Nos. 11722327, 11653002, 11961131007, 11725312, 11421303), by the CAST (2016QNRC001), by the National Youth Thousand Talents Program of China, and by the Fundamental Research Funds for Central Universities. 
The work of DAE is supported in part by the Foundational Questions Institute.
All numerics were operated on the computer clusters {\it LINDA \& JUDY} in the particle cosmology group at USTC. \\

\end{document}